\documentclass[aps,prx,twocolumn,floatfix,superscriptaddress]{revtex4-2}
\usepackage{graphicx,amsmath,amsfonts,mathrsfs,bbold,dsfont}
\begin{document}

\title{Ferromagnetic ferroelectricity due to the Kugel-Khomskii mechanism of the orbital ordering assisted by atomic Hund's second rule effects}

\author{I.~V.~Solovyev}
\email{SOLOVYEV.Igor@nims.go.jp}
\affiliation{Research Center for Materials Nanoarchitectonics (MANA), National Institute for Materials Science (NIMS), 1-1 Namiki, Tsukuba, Ibaraki 305-0044, Japan}
\author{R.~Ono}
\affiliation{Research Center for Materials Nanoarchitectonics (MANA), National Institute for Materials Science (NIMS), 1-1 Namiki, Tsukuba, Ibaraki 305-0044, Japan}
\author{S.~A.~Nikolaev}
\affiliation{Department of Materials Engineering Science, The University of Osaka, Toyonaka 560-8531, Japan}

\date{\today}

\date{\today}
\begin{abstract}
The exchange interactions in insulators depend on the orbital state of magnetic ions, obeying certain phenomenological principles, known as  Goodenough-Kanamori-Anderson rules. Particularly, the ferro order of alike orbitals tends to stabilize antiferromagnetic interactions, while the antiferro order of unlike orbitals favors ferromagnetic interactions. The Kugel-Khomskii theory provides a universal view on such coupling between spin and orbital degrees of freedom, based on the superexchange processes: namely, for a given magnetic order, the occupied orbitals tend to arrange in a way to further minimize the exchange energy. Then, if two magnetic sites are connected by the spatial inversion, the antiferro orbital order should lead to the ferromagnetic coupling \emph{and} break the inversion symmetry. This constitutes the basic idea of our work, which opens a new route for designing ferromagnetic ferroelectrics -- the rare but fundamentally and practically important multiferroic materials. After illustrating the basic idea on toy-model examples, we propose that such behavior can be indeed realized in the van der Waals ferromagnet VI$_3$, employing for this analysis the realistic model derived from first-principles calculations for magnetic $3d$ bands. We argue that the intraatomic Coulomb interactions responsible for Hund's second rule, acting against the crystal field, tend to restore the orbital degeneracy of the ionic $d^{2}$ state in VI$_3$ and, thus, provide a necessary flexibility for activating the Kugel-Khomskii mechanism of the orbital ordering. In the honeycomb lattice, this orbital ordering breaks the inversion symmetry, stabilizing the ferromagnetic-ferroelectric ground state. The symmetry breaking leads to the canting of magnetization, which can be further controlled by the magnetic field, producing a huge change of electric polarization. 
\end{abstract}

\maketitle

\section{\label{sec:Intro} Introduction}
\par In a broad sense, multiferroics are materials, where the ferroelectric (FE) order can coexist with a magnetic one~\cite{CheongMostovoy,Khomskii2009}. These are the key material systems for achieving the cross-control of magnetic and electric properties by applying an electric or magnetic field~\cite{TokuraSekiNagaosa}. Nevertheless, literally, the multiferroicity implies somewhat narrower requirement: both orders should be of the \emph{ferro} type, so that the material is not simply magnetic but ferromagnetic~\cite{Hill,Eerenstein,Tokura}. This is particularly important for the cross-control applications: if the ferromagnetic (FM) moment, $\boldsymbol{M}$, is finite and preferably large, it can be manipulated by a relatively weak magnetic field. The same holds for the ferroelectric polarization, $\boldsymbol{P}$, and the electric field. Thus, from practical point of view, it is desirable to have materials with large $\boldsymbol{M}$ and $\boldsymbol{P}$, and strong coupling between them. However, the ferroelectricity and ferromagnetism obeys very different principles and very rarely coexist in nature. Namely, the ferroelectricity implies breaking of the inversion symmetry. However, it cannot be achieved by a simple FM arrangement of spins, which has the same symmetry as the crystallographic one. On the other hand, if the inversion symmetry breaking results from the intrinsic instability of the crystal structure, there is no guarantee that the corresponding to it magnetic structure will be ferromagnetic. In fact, most of insulating transition-metal oxides are antiferromagnetic. 

\par Therefore, the main attention is paid to creation of artificial materials, which would combine the FE and FM characteristics within one sample or device~\cite{Eerenstein}. One possible direction is the synthesis of heterostructures, consisting of FE and FM layers of two different materials~\cite{heterostructures}. Another promising direction is the strain engineering. Particularly, some transition-metal oxides can turn into the FE-FM state by epitaxial strain~\cite{FennieRabe,EuTiO3Nature,LeeRabe}. The main driving force is the intrinsic FE instability of the so-called $d^{0}$ materials, related to the coupling between the occupied bonding and unoccupied antibonding states of opposite parity~\cite{Bersuker1966,Bersuker2012}. For instance, the coupling between the occupied O $2p$ and unoccupied Ti $3d$ bands in EuTiO$_3$, caused by the FE displacements, can low the energy~\cite{FennieRabe}. Moreover, the magnetic Eu$^{2+}$ ions alter this coupling, making it dependent on magnetic structure of the Eu sublattice. Thus, although cubic EuTiO$_3$ is the paraelectric antiferromagnet, the epitaxial strain can turn it into the FE-FM state~\cite{FennieRabe,EuTiO3Nature}. The partial occupation of antibonding transition-metal $3d$ states weakens the FE instability. Nevertheless, the effect can still persist for certain $3d$ configurations, such as $d^{3}$~\cite{Bersuker2012}, as was theoretically proposed for SrMnO$_3$~\cite{LeeRabe,EdstromEderer}, where the same Mn$^{3+}$ ions are responsible for magnetism and participate in the FE displacements, thus, resulting in stronger spin-lattice coupling and larger magnetic transition temperature in comparison with EuTiO$_3$. 

\par In this article we propose a completely new and so far unexplored route for designing ferroelectric (or polar) ferromagnets, which is based on the Kugel-Khomskii (KK) mechanism of the orbital ordering~\cite{KugelKhomskii}.

\par The interatomic exchange interactions between spins depend on the orbital state of atoms participating in these exchange processes: which orbitals are occupied, which are empty, and how they are oriented relatively to each other in the magnetic bonds, i.e. what is commonly called the orbital ordering~\cite{KugelKhomskii}. The basic rules describing character of these interactions in insulators are widely know as Goodenough-Kanamori-Anderson (GKA) rules~\cite{Anderson1950,Goodenough1955,Goodenough1958,Kanamori1959}. Particularly, the ferro orbital order, where electrons occupy the same orbitals, typically leads to the antiferromagnetic (AFM) coupling between the spins. On the other hand, the antiferro orbital order, where occupied orbitals alternate on the lattice, usually favors the FM interactions. These fundamental principles were further elaborated by Kugel and Khomskii (KK)~\cite{KugelKhomskii,KugelKhomskii1972,KugelKhomskii1973} on the basis of superexchange (SE) theory~\cite{Anderson1959}, resulting in what is now called the KK mechanism of the orbital ordering, which states that for a given spin order, the orbital degrees of freedom will tend to relax in the direction to further minimize the exchange energy. 

\par The KK mechanism was proposed long before the current era of multiferroic materials and so far has not been considered as a possible source of breaking the inversion symmetry. Typical applications of the KK mechanism are focused on the analysis of spin and orbital phenomena in compounds, where magnetic sites are located in the inversion centers and the materials remain centrosymmetric irrespectively of the spin or orbital order~\cite{KhomskiiStreltsov}, as in colossal magnetoresistive manganites~\cite{Maezono1998} or other perovskite transition-metal oxides~\cite{Mochizuki2003}. In fact, many of these materials do exhibit the antiferro orbital order, which is responsible for the FM character of exchange interactions, as it happens, for instance, in YTiO$_3$~\cite{Mochizuki2003}, LaMnO$_3$~\cite{Goodenough1955,KugelKhomskii}, or BiMnO$_3$~\cite{dosSantos,NJP}. However, the existence of inversion centers makes most of them anti-polar~\cite{Baettig}. 

\par What if the inversion center is located between two magnetic sites? Then, one can expect that the antiferro orbital order across the inversion center will lead to the FM interactions between the spins, as required by GKA rules, \emph{and} break the inversion symmetry, giving us a unique possibility for realizing simultaneously the ferromagnetism and ferroelectricity within one phase. This constitutes the main idea of our work, which will be elaborated as follows.

\par First, in Sec.~\ref{sec:model}, we will explore this basic idea by considering toy-model examples of degenerate $yz$ and $zx$ orbitals in the zigzag chain and honeycomb lattice, where the problem can be solved analytically providing a transparent expression for the exchange energy, which explains the emergence of the antiferro orbital order and electric polarization. The key aspect of the zigzag chain and honeycomb lattice is that both of them are centrosymmetric. However, the inversion centers are located in the mid-points connecting two magnetic sites. Therefore, these are the structures where the antiferro orbital order will simultaneously break the inversion symmetry and stabilize the FM ground state. 

\par
\begin{figure}[b]
\begin{center}
\includegraphics[width=8.6cm]{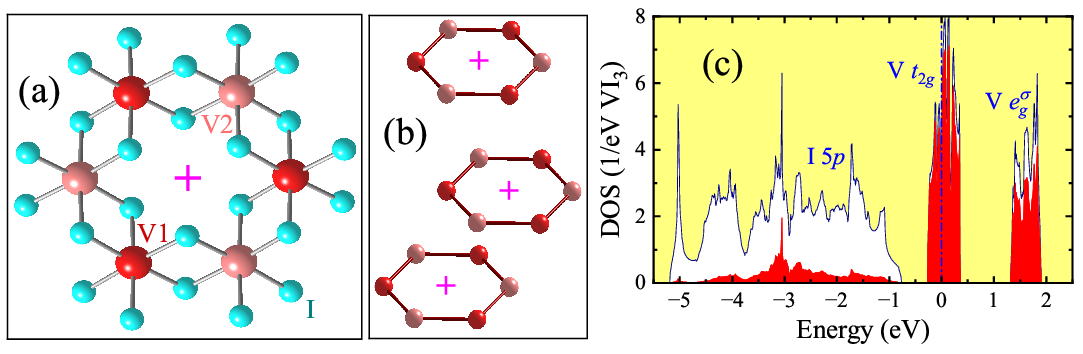} 
\end{center}
\caption{
(a) Fragment of the crystal structure of VI$_3$: each V atoms is surrounded by six I atoms, forming the hexagon of edge-sharing VI$_6$ octahedra. The vanadium sublattices, which are transformed to each other by the inversion operation are shown by different colors and denoted as V1 and V2. The inversion centers are denoted by $+$. (b) Stacking of the honeycomb planes. (c) Densities of states (DOS) in the local-density approximation. Shaded areas show partial contributions of the V $3d$ states. The Fermi level is at zero energy.}
\label{fig.LDA}
\end{figure}

 Then, in Sec.~\ref{sec:VI3}, we will turn to realistic example of VI$_3$, which has attracted a considerable attention as a new layered FM semiconductor with relatively high Curie temperature $T_{\rm C} \simeq 50$ K~\cite{VI3Kong}. The main structural motif of this quasi-two-dimensional van der Waals ferromagnet is again the honeycomb planes (Fig.~\ref{fig.LDA}). 
\noindent
\noindent According to formal valence arguments, each V site has two $3d$ electrons. In the octahedral environment they populate two out of three $t_{2g}$ orbitals, indicating the importance of orbital degrees of freedom in the physics of VI$_3$. The main question is, however, how well these orbital degrees of freedom are quenched by the local distortions of the VI$_6$ octahedra. Indeed, the distortions will tend to split the $t_{2g}$ levels. The fundamental Jahn-Teller theorem states in this respect that the splitting should lift the orbital degeneracy in the direction to form a nondegenerate ground state~\cite{JahnTeller}. Nevertheless, if the splitting is small, other ingredients can come into play. Particularly, two electrons in the $3d$ shell are subjected to Hund's rule effects, which act in the opposite direction and tend to reenforce the ground state with maximal multiplicity. The corresponding energy gain is controlled by the Racah parameter $B$~\cite{Racah,Slater}. Using electronic structure calculations based on density functional theory (DFT), we will evaluate relevant parameters and show that $B$ in VI$_3$ is sufficiently large to overcome the crystal-field splitting and activate the KK mechanism of the orbital ordering, as it will follow from the analysis of atomic multiplet structure and dynamical mean-filed theory (DMFT) calculations on the honeycomb lattice~\cite{DMFT1,DMFT2}. Then, we will show that for the realistic parameters range, the antiferro orbital order can be indeed established in VI$_3$, resulting in the FM-FE ground state. The relativistic spin-orbit (SO) interaction interplays with the symmetry breaking caused by the orbital ordering, resulting in a canted magnetic structure, which can be further controlled by the magnetic field, leading to a huge change of electric polarization. 

\par Finally, in Sec.~\ref{sec:Summary}, we will summarize our results, discussing their implications to the properties of VI$_3$ as well as more general aspects of the Hund's rule physics in solids.

\section{\label{sec:model} Toy-model considerations}
\par The goal of this section is to illustrate the basic idea of inversion symmetry breaking by the orbital ordering, resulting in coexistence of ferroelectricity and ferromagnetism. For these purposes we consider toy-model examples of degenerate $yz$ and $zx$ orbitals in the zigzag chain and honeycomb lattice. 

\subsection{\label{sec:zigzag} Ordering of the $yz$ and $zx$ orbitals in the zigzag chain}
\par
\begin{figure}[t]
\begin{center}
\includegraphics[width=8.6cm]{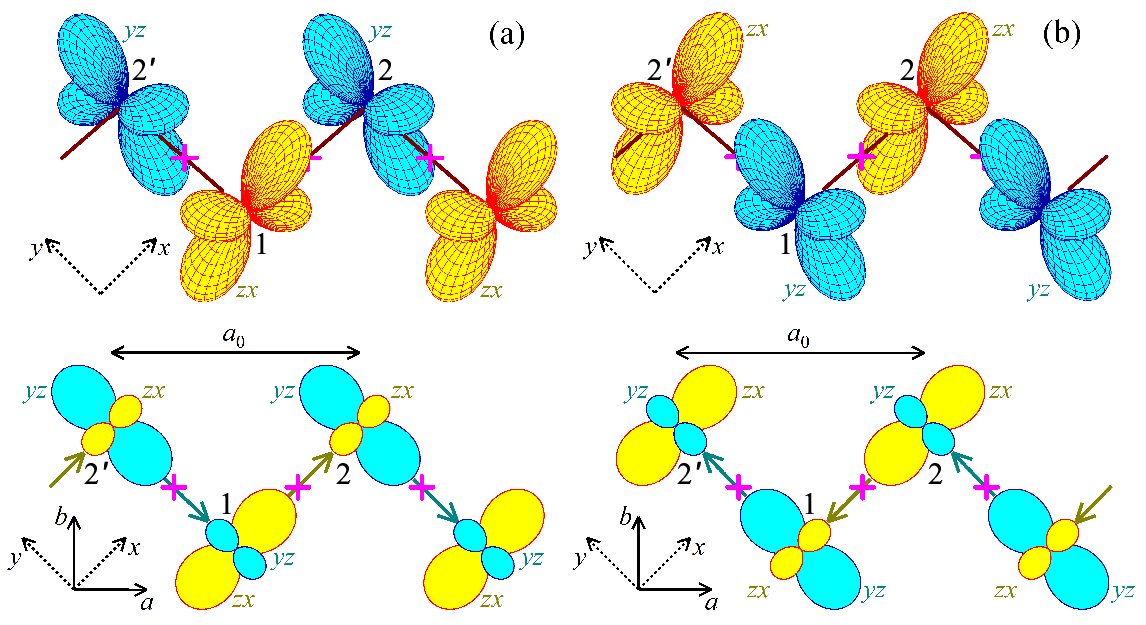} 
\end{center}
\caption{
Ordering of the $yz$ and $zx$ orbitals in the zigzag chain, breaking the inversion symmetry and stabilizing the ferromagnetic coupling: side view (upper panel) and top view (lower panel). The electron densities across the inversion centers (denoted by $+$) are plotted by different colors: larger objects are the densities in the atomic limit and smaller objects are the densities transferred from the neighboring sites due to the superexchange processes in the directions, which are shown by arrows. $a_{0}$ is the lattice parameter.}
\label{fig.zigzag}
\end{figure}

 The simplest model, which explains the basic physics of how the KK mechanism can break the inversion symmetry and induce the electric polarization is the one-dimensional zigzag chain (see Fig.~\ref{fig.zigzag}). In this case, there are two sites in the unit cell ($1$ and $2$), which can be transformed to each other by the spatial inversion about the mid-point of the bond, connecting these two sites. Let us assume that there is only one electron per site, which is shared by two atomic states, $yz$ and $zx$. Thus, in the atomic limit, the ground state is degenerate. Then, the electron hoppings $\hat{t}_{ij}$ are such that in the neighboring bonds they will connect $zx$ with $zx$ in the direction $x$ and $yz$ with $yz$ in the direction $y$~\cite{SK}: $t^{zx,zx}_{ij || x} = t^{yz,yz}_{ij || y} = t$. This hopping lifts the degeneracy, ordering the orbitals in the alternating way, as explained in Fig.~\ref{fig.zigzag}, which minimizes the energy of SE interactions for the FM state~\cite{KugelKhomskii}.
\noindent
\noindent The same orbital ordering makes the atomic sites inequivalent and, thus, breaks the inversion symmetry. The corresponding electric polarization can be evaluated along the same line as in the theory of SE interactions~\cite{PRB2013,PRB2020} but starting for these purposes with the general expression for $\boldsymbol{P}$ in periodic systems, formulated in terms of the Wannier functions~\cite{FE_theory1,FE_theory2,FE_theory3}. Namely, if $| \alpha_{i}^{o} \rangle$ is the occupied Wannier function at site $i$ in atomic limit, $\hat{t}$ will induce the tail of this orbital, $| \alpha_{i \to j}^{o} \rangle$, spreading to the neighboring site $j$. It can be evaluated by treating $\hat{t}$ as a perturbation, the same as in the SE theory, which yields $| \alpha_{i \to j}^{o} \rangle = -\frac{1}{\Delta} | \alpha_{j}^{u} \rangle \langle \alpha_{j}^{u} | \hat{t}_{ji} | \alpha_{i}^{o} \rangle $, where $\Delta$ is an effective on-site interaction describing the splitting of occupied and unoccupied states and $| \alpha_{j}^{u} \rangle$ is the unoccupied Wannier function. This yields $\boldsymbol{P} || \boldsymbol{a} = \mp e \left( \frac{t}{\Delta} \right)^2$~\cite{SM}, where two signs stand for the orbital ordering depicted in Figs.~\ref{fig.zigzag}a ($+$) and \ref{fig.zigzag}b ($-$), and $e$ is the minus electron charge. In the one-dimensional case, $\boldsymbol{P} || \boldsymbol{a}$ is nothing but the edge charge~\cite{FE_theory2}.

\par Similar model was considered in Refs.~\cite{Barone,PRB2013} to explain the emergence of electric polarization in the E-phase of manganites. The main difference is that, in manganites, the orbital order is driven by the Jahn-Teller distortion, which is an external factor in the considered electronic model, while here it originates solely from the SE interactions and, formally, no distortion is needed to break the inversion symmetry. 

\subsection{\label{sec:honeycomb} Ordering of the $yz$ and $zx$ orbitals in the honeycomb plane}
\par
\begin{figure}[t]
\begin{center}
\includegraphics[width=4.2cm]{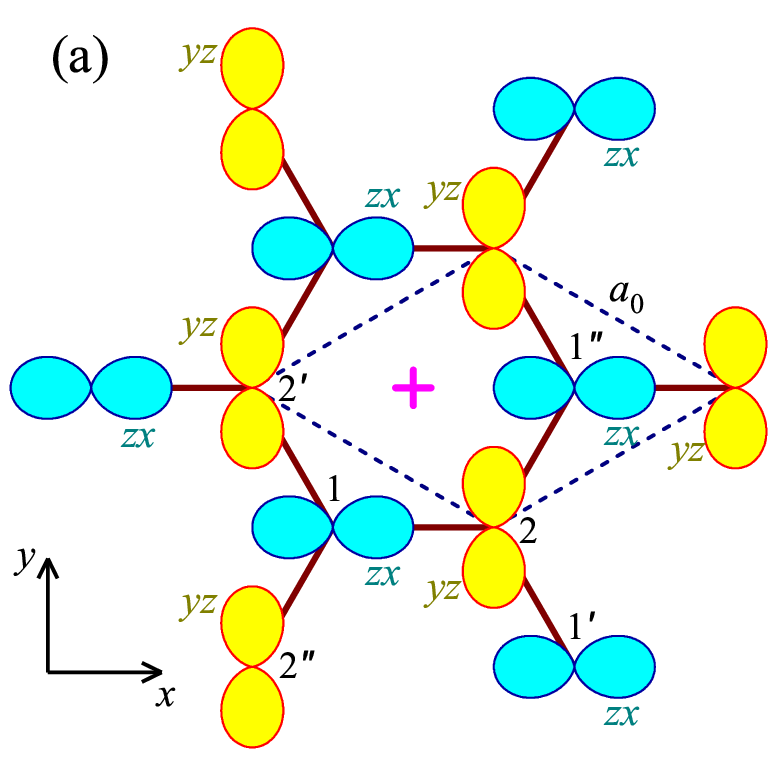} \includegraphics[width=4.2cm]{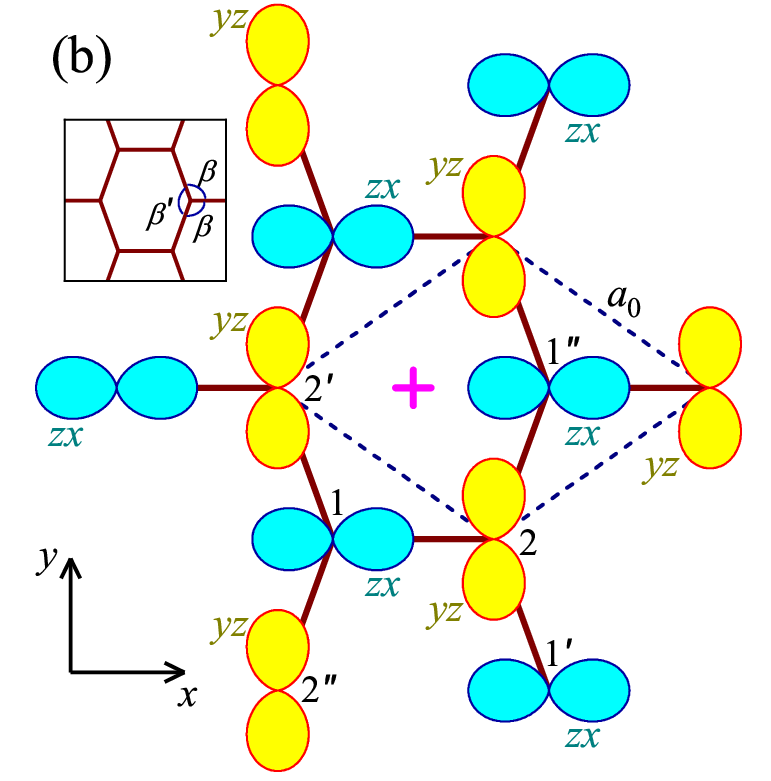} 
\end{center}
\caption{
Ordering of $yz$ and $zx$ orbitals in the honeycomb plane, breaking the inversion symmetry and stabilizing ferromagnetic interactions. The electron densities across the inversion centers (denoted by $+$) are plotted by different colors. The unit cell is shown by dashed line. $a_{0}$ is the lattice parameter. (a) Ideal lattice. (b) Distorted lattice, where $\beta \ne \beta'$ (as explained in the inset).}
\label{fig.honeycomb}
\end{figure}

 Now we turn to a more realistic model of the honeycomb plane, which may have some relevance to realistic materials, such as TiCl$_3$~\cite{TiCl3}. In the honeycomb lattice, there are also two sites in the unit cell, which can be transformed to each other by the spatial inversion (see Fig.~\ref{fig.honeycomb}a). Again, we assume that there are only two orbitals, $yz$ and $zx$, and one electron per site. The transfer integral operates only between orbitals, which are parallel to the bond. For instance, for the bond $1$-$2$ in Fig.~\ref{fig.honeycomb}, these are $zx$ orbitals. The transfer integrals in other bonds can be obtained by threefold rotations, as explained in Supplemental Material~\cite{SM}.
\noindent
\noindent

\par Defining at each site ($\nu$) of the unit cell the occupied ($o$) and unoccupied ($u$) orbitals as 
\noindent
\begin{eqnarray*}
| \alpha^{o}_{\nu} \rangle & = & \phantom{-} \cos \phi_{\nu} | yz \rangle + \sin \phi_{\nu} | xz \rangle \\
| \alpha^{u}_{\nu} \rangle & = & -           \sin \phi_{\nu} | yz \rangle + \cos \phi_{\nu} | xz \rangle,
\end{eqnarray*}
\noindent and treating the transfer integrals as a perturbation, it is straightforward to find the following expression for the total energy change~\cite{SM}:
\noindent
\begin{eqnarray*}
{\cal E} = -\frac{3t^2}{4\Delta} \bigg( 2 - \cos2(\phi_{1} - \phi_{2}) \bigg),
\end{eqnarray*}
\noindent which takes the minimum if $\phi_{2} = \phi_{1} + \frac{\pi}{2} \mod \pi$. Thus, one of the angles, $\phi_{1}$, remain unspecified. Then, the polarization is given by:
\begin{eqnarray*}
\boldsymbol{P} \equiv \bigg( \begin{array}{c} P_{x} \\ P_{y} \end{array} \bigg)
= \frac{e}{a_{0}} \left( \frac{t}{\Delta} \right)^2
\bigg(
\begin{array}{r}
-\cos 2\phi_{1} \\
\sin 2\phi_{1} 
\end{array}
\bigg),
\end{eqnarray*}
\noindent where $a_{0}$ is the lattice parameter. Thus, the orbital ordering breaks the spatial inversion, yielding finite $|\boldsymbol{P}| = \frac{e}{a_{0}} \big( \frac{t}{\Delta} \big)^2$. However, the ground state remains degenerate and $\boldsymbol{P}$ can have any direction in the $xy$ plane, depending on the angle $\phi_{1}$.

\par There can be several scenarios of lifting the degeneracy. For instance, in a more general case, where there are several unoccupied states, such degeneracy does not occur (see Sec.~\ref{sec:oo}). In the simples two-orbital model, considered here, the value of $\phi_{1}$ can be decided by the exchange striction effects. Particularly, the orbital ordering in Fig.~\ref{fig.honeycomb}a will make the bond $1$-$2'$ and $1$-$2''$ different from the bond $1$-$2$ and the structure will tend to relax in order to further minimize the energy change. Here, we assume that such deformation of the honeycomb plane can be described by the angle $\beta = \frac{2\pi}{3} + \delta \beta$, formed by the bonds $1$-$2'$ and $1$-$2''$ with the bonds $1$-$2$, which is different from the angle $\beta' = \frac{2\pi}{3} -2 \delta \beta$, formed by the bonds $1$-$2'$ and $1$-$2''$ with each other, while the bondlengths are assumed to be the same. $\delta \beta = 0$ corresponds to the undistorted structure. The situation is explained in Fig.~\ref{fig.honeycomb}b.

\par The transfer integrals in the bond $1$-$2$ operate only between orbitals $zx$, while the ones in the bonds $1$-$2'$ and $1$-$2''$ are obtained by considering rotations of $1$-$2$ by the angle $\mp \beta$~\cite{SM}. For small $\delta \beta$, the corresponding energy change is given by 
\noindent
\begin{eqnarray*}
{\cal E} = -\frac{3t^2}{4\Delta} \left( 2 - \cos 2(\phi_{1} - \phi_{2}) + \frac{4}{\sqrt{3}} \delta \beta \cos 2(\phi_{1} + \phi_{2}) \right).
\end{eqnarray*}
\noindent Then, $\delta \beta > 0$ strengthens the antiferro orbital order in the bonds $1$-$2'$ and $1$-$2''$. In this case, the 2nd and 3rd terms in $\left( \dots \right)$ can be minimized independently, yielding $\phi_{1} = 0$ and $\phi_{2} = \frac{\pi}{2}$. Thus, the degeneracy is lifted and the polarization is parallel to the $x$ axis.

\par The considered models dealing with the $d^{1}$ systems implies that the degenerate $yz$ and $zx$ states are split off by the crystal field to become the lowest energy atomic states, which accommodate single electron. Although these models are easy to solve, and in this sense can be very insightful, they are hardly practical. The main obstacle for the practical realization of the considered scenarios is the direction of the crystal field, which typically act to form a nondegenerate ground state, as it is required by the Jahn-Teller theorem~\cite{JahnTeller}. For instance, one possible $d^{1}$ candidate to form the antiferro orbital order in the honeycomb lattice is TiCl$_3$~\cite{TiCl3}. However, the crystal field in TiCl$_3$ tend to stabilize the nondegenerate $z^{2}$ orbital, in the direction perpendicular to the honeycomb plane, while the degenerate states lie higher in energy. The basic limitation of the $d^{1}$ system is that the crystal field is the only parameter, which can control the order of the atomic states. Due to the one-electron character of the problem, the atomic Hund's rules simply not apply here. Thus, there is no way to reverse the order of the crystal-field orbitals in the favor of degenerate ground state. In this sense, a more promising direction is to explore $d^{2}$ materials. Then, the crystal field will still tend to select a nondegenerate ground state. Nevertheless, the new aspect of the $d^{2}$ systems is that the interaction between two $3d$ electrons is subjected to the Hund's rule effects, which act in the opposite direction and tend to stabilize the ground state with maximal orbital multiplicity. In the next section, we will argue that VI$_3$ is indeed a good candidate for practical realization of such scenario.


\section{\label{sec:VI3} Implications to the properties of VI$_3$}
\par VI$_3$ exhibits the structural phase transition at $T_{\rm s} \simeq 78$ K, which is followed by the FM transition at $T_{\rm C} \simeq 50$ K~\cite{VI3Kong}. Another structural phase transition, at around $32$ K, was also suggested~\cite{VI3Dolezal,structure}. However, currently there is no clear consensus even about the crystallographic symmetry of VI$_3$. For instance, the high-temperature ($T > T_{\rm s}$) phase was proposed to have trigonal $R\overline{3}$ (space group No. 148)~\cite{VI3Kong,VI3Dolezal,structure}, trigonal $P\overline{3}1c$ (163)~\cite{VI3Son}, and monoclinic $C2/m$ (12)~\cite{VI3Tian} symmetry. The low-temperature ($T < T_{\rm s}$) phase was proposed to be triclinic $P\overline{1}$ (2)~\cite{structure,VI3Hao}, monoclinic $C2/c$ (15)~\cite{VI3Son}, and trigonal $R\overline{3}$ (148)~\cite{VI3Tian}. Thus, even the direction of the distortion with lowering temperatures appears to be the subject of controversy: some reports suggest symmetry lowering~\cite{structure,VI3Son}, while another report suggests that the symmetry becomes higher, similar to what is observed in CrI$_3$~\cite{VI3Tian}. To a certain extent, the value of $T_{\rm s}$ can be controlled by the magnetic field~\cite{VI3Dolezal}. Moreover, the $32$ K transition was also suggested to be magnetic, due to either disappearance of the magnetic order in one of the V sublattices~\cite{VI3Gati} or reorientation of the magnetic moments~\cite{VI3Hao}.

\par In this section, we systematically study the symmetry breaking in VI$_3$ caused by the orbital ordering, starting for these purposes with the structure with highest $R\overline{3}$ symmetry~\cite{structure}. The $R\overline{3}$ space group can be generated by considering the threefold rotations about $z$ in combination with spatial inversion. We will show that the experimentally observed breaking of the threefold rotation symmetry can be rationalized by considering the KK mechanism of the orbital ordering in combination with the Hund's second rule effects. Furthermore, we predict the new symmetry pattern, also driven by the KK mechanism, where both threefold rotation and inversion symmetries are broken by the orbital ordering. The prediction remains largely intact even for the low-symmetry experimental $P\overline{1}$ structure. This structure has inversion symmetry. However, it can be again broken by the orbital order. We evaluate the electric polarization, induced by the inversion symmetry breaking, and propose how it can be controlled by magnetic field.

\subsection{\label{sec:method} Method}
\par The electronic structure calculations have been performed for the experimental crystal structure of VI$_3$, reported in Ref.~\cite{structure}, and using for these purposes either linear muffin-tin orbital (LMTO) method~\cite{LMTO1,LMTO2} or pseudopotential Quantum ESPRESSO (QE) method~\cite{QuantumE}, which were supplemented with, respectively, local density approximation (LDA)~\cite{VWN} and generalized gradient approximation (GGA)~\cite{PBE} for the exchange-correlation potential. After that, we construct the 5-orbital model for the magnetic V $3d$ bands located near the Fermi level (V$t_{2g}$$+$V$e_{g}$ bands in Fig.~\ref{fig.LDA}c):
\noindent
\begin{widetext}
\begin{equation}
\hat{\cal{H}}  =  \sum_{ij} \sum_{\sigma \sigma'} \sum_{ab}
t_{ij}^{a \sigma, b \sigma'}
\hat{c}^\dagger_{i a \sigma}
\hat{c}^{\phantom{\dagger}}_{j b \sigma'} +
  \frac{1}{2}
\sum_{i}  \sum_{\sigma \sigma'} \sum_{abcd} U^{abcd}
\hat{c}^\dagger_{i a \sigma} \hat{c}^\dagger_{i c \sigma'}
\hat{c}^{\phantom{\dagger}}_{i b \sigma}
\hat{c}^{\phantom{\dagger}}_{i d \sigma'},
\label{eqn.ManyBodyH}
\end{equation}
\end{widetext}
\noindent where $\hat{c}^\dagger_{i a \sigma}$ ($\hat{c}_{i a \sigma}$) stands for the creation (annihilation) of an electron with spin $\sigma$ at the Wannier orbital $a$ of site $i$. For the construction of Wannier functions in QE, we use the maximally localized Wannier functions method~\cite{WannierRevModPhys}, as implemented in the wannier90 package~\cite{wannier90}, while in LMTO we employ the projector-operator technique~\cite{WannierRevModPhys,review2008}. Then, the one-electron part in (\ref{eqn.ManyBodyH}), $\hat{t} = [t_{ij}^{a \sigma, b \sigma'}]$, is given by the matrix elements of the LDA (GGA) Hamiltonian in the Wannier basis. The dependence of $t_{ij}^{a \sigma, b \sigma'}$ on the spin indices is due to the SO interaction, where the main contributions come from the heavy I sites. In order to include these contributions, it is essential that the electronic structure for the V $3d$ bands should be calculated with the SO coupling \emph{before} the construction of the model Hamiltonian. Without SO coupling, the matrix elements $t_{ij}^{a \sigma, b \sigma'}$ become  $t_{ij}^{a \sigma, b \sigma'} = t_{ij}^{a, b} \delta_{\sigma \sigma'}$.

\par The screened Coulomb interactions are evaluated within constrained random-phase approximation~\cite{cRPA}. In LMTO, this is done in an approximate way, basically by considering the self-screening effect of on-site Coulomb interactions in the V $3d$ bands by the same V $3d$ states, which are admixed into the I $5p$ and other bands (see Fig.~\ref{fig.LDA}c), as explained in Ref.~\cite{review2008}. After that, the $5$$\times$$5$$\times$$5$$\times$$5$ matrix $\hat{U} = [U^{abcd}]$ of screened intra-atomic Coulomb interactions was fitted in terms of three parameters, which would describe these interactions in spherical atomic environment: the Coulomb repulsion $U = F^0$, responsible for the overall stability of atomic shell with the given number of electrons; the intra-atomic exchange coupling $J=(F^2$$+$$F^4)/14$, responsible for Hund's first rule; and the Racah parameter $B = (9F^2$$-$$F^4)/441$, responsible for Hund's second rule (where $F^0$, $F^2$, and $F^4$ are Slater integrals)~\cite{Racah,Slater}. The obtained parameters $U$, $J$, and $B$ are listed in Table~\ref{tab:Coulomb}.
\noindent
\begin{table}[h]
\caption{Parameters $U$, $J$, and $B$ of intra-atomic Coulomb interactions, and trigonal splitting $\Delta_{\rm tr}$ between $e_{g}^{\pi}$ and $a_{1g}^{\phantom{\pi,2}}$ levels (all are in eV) in the $R\overline{3}$ phase of VI$_3$, as obtained in LMTO and QE methods.}
\label{tab:Coulomb}
\begin{ruledtabular}
\begin{tabular}{ccccc}
method & $U$     & $J$    & $B$    & $\Delta_{\rm tr}$ \\
\hline
LMTO   & $1.21$  & $0.74$ & $0.07$ & $0.01$            \\
QE     & $1.97$  & $0.64$ & $0.06$ & $0.03$
\end{tabular}
\end{ruledtabular}
\end{table}
\noindent The value of $U$ is somewhat smaller in LMTO, while the values of $J$ and $B$, obtained LMTO and QE methods, are comparable. 

\subsection{\label{sec:Hund} Hund's second rule and orbital degeneracy}
\par
\begin{figure*}[t]
\begin{center}
\includegraphics[width=17.0cm]{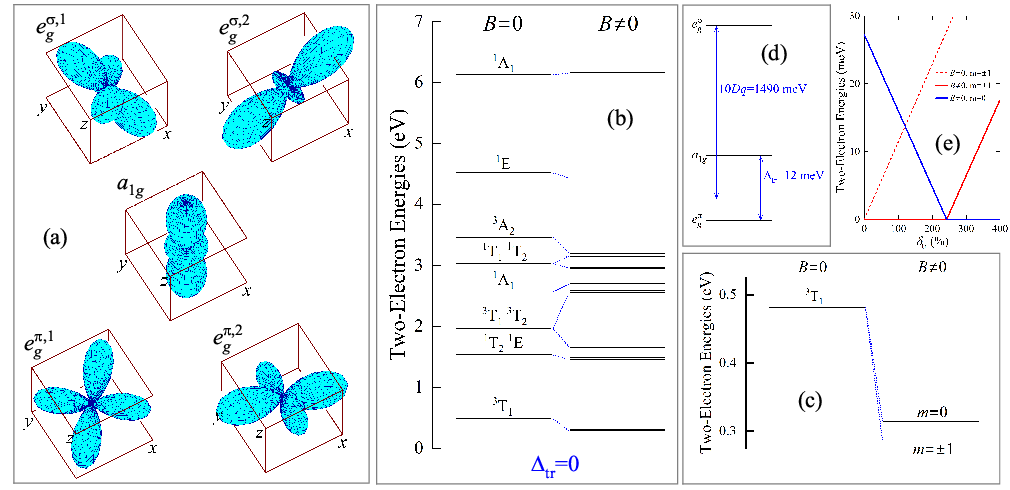} 
\end{center}
\caption{
(a) Crystal-field orbitals for VI$_3$. (b) The effect of Racah parameter $B$ on the two-electron energies without trigonal crystal field. (c) Amplified low-energy part of (b). (d) Parameters of the crystal field splitting and their values obtained in the LMTO method. (e) Two-electron energies, relative to the lowest state, depending on the trigonal splitting $\delta_{\rm tr} = 100 (\Delta_{\rm tr}/\Delta_{\rm tr}^0)$, where $\Delta_{\rm tr}^0 = 12$ meV is the actual value obtained in the LMTO calculations. The value of $B$ is set to 70 meV.}
\label{fig.atomic}
\end{figure*}

 In this section we consider interplay between crystal field and Coulomb interactions in the atomic limit. The crystal-field orbitals, obtained after the diagonalization of the site-diagonal part of $\hat{t}$ are shown in Fig.~\ref{fig.atomic}a.
\noindent
\noindent The octahedral field, $10Dq$, splits the $3d$ levels into the three-dimensional representation $t_{2g}$ and two-dimensional representation $e_{g}^{\sigma}$, 
\noindent
\begin{eqnarray*}
| e_{g}^{\sigma,1} \rangle & = & -\sin \alpha | xy      \rangle + \cos \alpha | zx \rangle \\
| e_{g}^{\sigma,2} \rangle & = & -\sin \alpha | x^2-y^2 \rangle + \cos \alpha | yz \rangle.
\end{eqnarray*}
\noindent The $t_{2g}$ levels are further split by the trigonal field $\Delta_{\rm tr}$ into the doubly-degenerate states $e_{g}^{\pi}$,
\noindent
\begin{eqnarray*}
| e_{g}^{\pi,1} \rangle & = & \cos \alpha | xy      \rangle + \sin \alpha | zx \rangle  \\
| e_{g}^{\pi,2} \rangle & = & \cos \alpha | x^2-y^2 \rangle + \sin \alpha | yz \rangle,
\end{eqnarray*}
\noindent which belong to the same representation as $e_{g}^{\sigma}$, and the nondegenerate state $| a_{1g}^{\phantom{\pi,2}} \rangle = | z^2 \rangle$. The numerical value of $\alpha$ is about $34^{\circ}$. Since $\Delta_{\rm tr} > 0$, the $e_{g}^{\pi}$ states lie lower than  $a_{1g}^{\phantom{\pi,2}}$ (see Fig.~\ref{fig.atomic}d) and accommodate both $3d$ electrons. Thus, from the viewpoint of the one-electron crystal-field splitting, the ground state is expected to be nondegenerate (the so-called $e_{g}^{\pi}e_{g}^{\pi}$ state~\cite{Ezhov}), in agreement with the Jahn-Teller theorem~\cite{JahnTeller}, which should be satisfied at the level of LDA/GGA calculations. Nevertheless, LDA (GGA) is an approximation, which does not properly take into account the interactions responsible for Hund's second rule~\cite{Weinert}. The letter are proportional to the Racah parameter $B$ and can easily overcome the trigonal splitting if $B > \Delta_{\rm tr}$. Realistic estimate of the parameters suggests that this situation is indeed realized in VI$_3$: $B$ is certainly small. However, $\Delta_{\rm tr}$ appears to be even smaller (see Table~\ref{tab:Coulomb}). This trend appears to be generic as very similar behavior was obtained for the triclinic modifications of VI$_3$ at $T$$=$ $9$ and $60$ K~\cite{structure}: the triclinic distortion further splits the $t_{2g}$ levels. However, the splitting ($\sim 10$ - $20$ meV~\cite{SM}) remains to be smaller than $B$.

\par Thus, the intra-atomic Coulomb interactions will tend to reverse the order of $e_{g}^{\pi}$ and $a_{1g}^{\phantom{\pi,2}}$ levels. This is due to the fundamental property of the Hund's rule coupling to form a ground state with the greatest value of multiplicity. The intuitive reason can be understood as follow: The main contribution to the $e_{g}^{\pi}$ states are associated with the $xy$ and $x^2$$-$$y^2$ orbitals (the corresponding $\cos \alpha \sim 0.83$), which are both located in the $xy$ plane and, therefore, experience a strong Coulomb repulsion. In order to reduce this repulsion, it is energetically more favorable to replace one of the occupied $e_{g}^{\pi}$ states by the $a_{1g}^{\phantom{\pi,2}}$ state. Simple mean-field considerations can be found in Supplemental Material~\cite{SM}, which clearly show that the effect is indeed driven by the Racah parameter $B$. 

\par Results of diagonalization of the two-electron Hamiltonian, combining the on-site Coulomb interactions with the crystal field are summarized in Fig.~\ref{fig.atomic}. First, we enforce $\Delta_{\rm tr} = 0$. Then, without $B$, the energy diagram is pretty much similar to the one obtained by Tanabe and Sugano for the $d^{2}$ configuration in octahedral environment~\cite{TanabeSugano} (Fig.~\ref{fig.atomic}b). Namely, the ground state is $^3$T$_1$, which is ninefold degenerate (being both spin and orbital triplet). Some states, like T$_1$ and T$_2$ remain accidentally degenerate when $B=0$. Finite $B$ lifts this degeneracy not only between T$_1$ and T$_2$, but also within each of these state. Particularly, the $^3$T$_1$ ground state is split into the sixfold degenerate state -- the spin triplet with orbital magnetic quantum numbers $m= \pm 1$ and threefold degenerate state spin triplet with $m=0$ (Fig.~\ref{fig.atomic}b). Then, we consider the effect of $\Delta_{\rm tr}$ (Fig.~\ref{fig.atomic}e): As expected, for small $\Delta_{\rm tr}$, the states with $m= \pm 1$ remains lower in energy, while further increase of $\Delta_{\rm tr}$ will eventually change the order of the $m= \pm 1$ and $m=0$ states. 

\par Thus, within realistic parameters range, the ground state of the V$^{3+}$ ion in VI$_3$ remains to be orbitally degenerate, opening a possibility for the KK mechanism of orbital ordering, once we consider the transfer integrals in the honeycomb lattice. 

\par It is also instructive to compare the behavior of VI$_3$ with V$_2$O$_3$. The latter compound, formally hosting the same V$^{3+}$ ions in the trigonal environment, was regarded as the canonical $S=\frac{1}{2}$ Mott insulators, where $a_{1g}^{\phantom{\pi,2}}$ electrons are dimerized in the V-V bonds and form a singlet, while the remaining $e_{g}^{\pi}$ electrons form an orbitally ordered state, resulting in a peculiar AFM structure~\cite{Castellani}. However, this picture was later revisited on the basis of first-principles calculations, suggesting the predominantly $e_{g}^{\pi} e_{g}^{\pi}$ configuration of V$^{3+}$ in V$_2$O$_3$, corresponding to the nondegenerate $m=0$ state~\cite{Ezhov}. Then, why the behavior of V$^{3+}$ in VI$_3$ can be so different from V$_2$O$_3$? We have constructed the model for the V $3d$ bands in V$_2$O$_3$ using the experimental $R\overline{3}c$ structure at $T = 175$ K~\cite{V2O3exp}. The corresponding parameters, controlling the order of the $m= \pm 1$ and $m=0$ states, are $B = 0.09$ eV and $\Delta_{\rm tr} = 0.15$ eV. Thus, although $B > \Delta_{\rm tr}$ in VI$_3$, it appears that $B < \Delta_{\rm tr}$ in V$_2$O$_3$, reversing the order of states and forcing the $m=0$ ground state. Furthermore, $10Dq$ is another important parameter in the problem. The Hund's rule interactions tend to mix the $e_{g}^{\pi}$ and $e_{g}^{\sigma}$ states, belonging to the same representation, to further minimize the energy of intra-atomic Coulomb interactions. This mixing acts against the octahedral field. Thus, the perspectives of realization of the orbitally degenerate ground state depend not only on the values of $B$ and $\Delta_{\rm tr}$, but also on $10Dq$. In this respect, it is also important that $10Dq$ is relatively small in VI$_3$ ($10Dq = 1.49$ eV) in comparison with V$_2$O$_3$ ($10Dq = 2.31$ eV). 

\par It is needless to say that the use of the 5-orbital model, constructed for the both $t_{2g}^{\phantom{\pi}}$ and $e_{g}^{\sigma}$ bands, is essential to incorporate the effects of Hund's second rule. Formally, the V$t_{2g}^{\phantom{\pi}}$ bands in VI$_3$ are well separated from other bands (see Fig.~\ref{fig.LDA}). Then, it would be straightforward to construct a more compact 3-orbital model for the V$t_{2g}^{\phantom{\pi}}$ bands, which is certainly easier to solve. However, there can be no separate Hund's second rule in the isolated $t_{2g}^{\phantom{\pi}}$ shell, where two-electron configuration with $S=1$ is equivalent to a non-interacting single-hole configuration. Therefore, such model would be meaningless for the purposes of our work, as it does not take into account the essential piece of physics. 

\subsection{\label{sec:el} Basic electronic structure}
\par
\begin{figure}[b]
\begin{center}
\includegraphics[width=8.6cm]{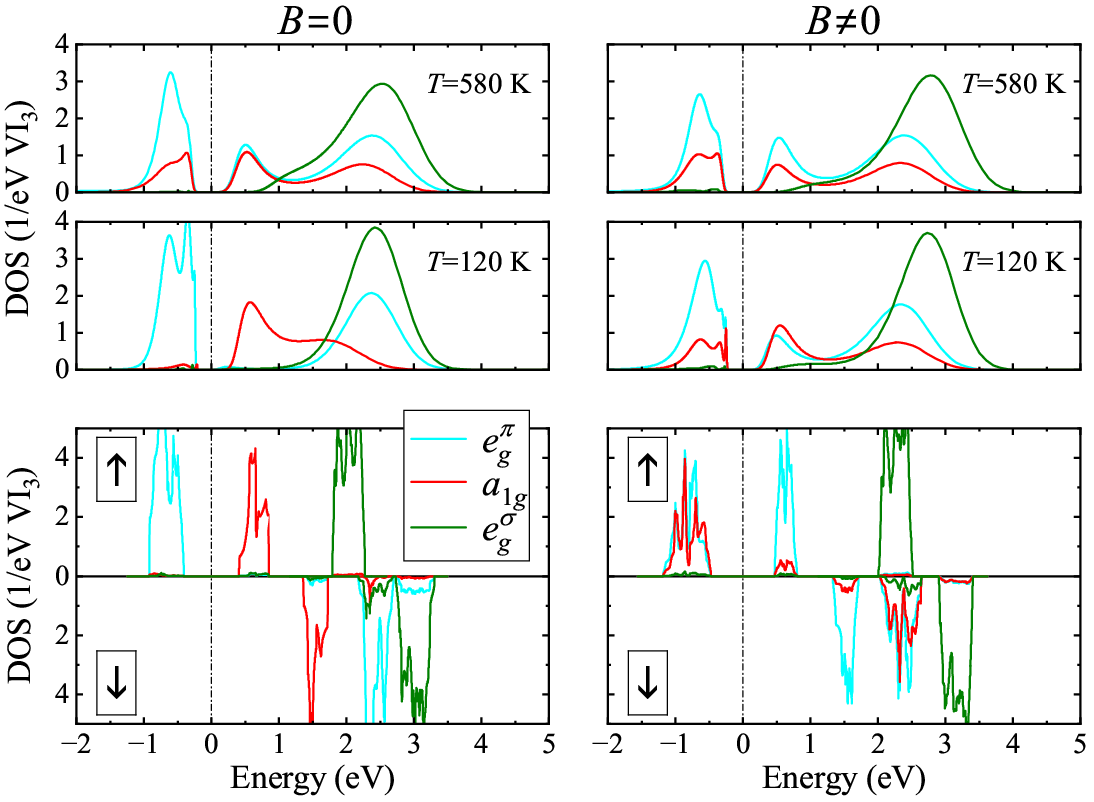} 
\end{center}
\caption{
Total and partial densities of states in the crystal-field representation as obtained in the dynamical mean-field theory for $T=$ $580$ and $120$ K in the paramagnetic state (top) and Hartree-Fock approximation for the ferromagnetic state (bottom). The Racah parameter $B$ was set to either $0$ (left) or $0.06$ eV (right). Other parameters were taken from the QE set (see Table~\ref{tab:Coulomb}). The Fermi level is at zero energy (the middle of the band gap).}
\label{fig.dos}
\end{figure}

 In most of the applications, the model (\ref{eqn.ManyBodyH}) was solved in mean-field Hartree-Fock (HF) approximation~\cite{review2008}, replacing the interaction part by
\noindent
\begin{equation}
\sum_{i} \sum_{\sigma  \sigma'} \sum_{ab} {\cal V}_{i,ab}^{\sigma \sigma'} \hat{c}^{ \sigma \dagger}_{i a} \hat{c}^{\sigma \phantom{\dagger}}_{i b}
\label{eqn.HF}
\end{equation}
\noindent and solving the obtained one-electron equations with the mean-field potential $\hat{\cal V}_{i} = [{\cal V}_{i,ab}^{\sigma \sigma'}]$ self-consistently. In order to check the validity of the HF method, we also employ the dynamical mean-field theory (DMFT). The details can be found in Supplemental Material~\cite{SM}. In order to reproduce the semiconducting behavior of VI$_3$ in the framework of DMFT, it is important to use the QE model parameters. Smaller value of $U$, obtained in the LMTO method, was insufficient to open the gap. The corresponding densities of states are shown in Fig.~\ref{fig.dos}.
\noindent

\par First, the HF method captures the main tendencies of two-electron calculations in the atomic limit considered in the previous section. Namely, without $B$, the doubly degenerate majority-spin $e_{g}^{\pi}$ states are occupied, while the nondegenerate $a_{1g}^{\phantom{\pi}}$ states are located in the unoccupied part of the spectrum, which is totally consistent with the fact that for $B=0$ the occupation of atomic states is controlled solely by the crystal field. Nevertheless, the situation changes dramatically when we switch on $B$. Then, the degeneracy of the $e_{g}^{\pi}$ states is lifted, splitting them into the occupied and unoccupied bands. Furthermore, the $a_{1g}^{\phantom{\pi}}$ states become occupied. This tendency is supported by DMFT calculations for the paramagnetic state. However, these calculations are performed at finite temperatures, which additionally affect the distribution of the atomic states. Particularly, $T=$ $580$, $230$, and $120$ K used in the calculations (all are far above $T_{\rm C}$) correspond to $k_{\rm B}T=$ $0.05$, $0.02$, and $0.01$ eV. The results for $T=$ $580$ and $120$ K are shown in Fig.~\ref{fig.dos} and the ones for $T=$ $230$ K are discussed in Supplemental Material~\cite{SM}. The first value of $k_{\rm B}T$ is larger than $\Delta_{\rm tr}= 0.03$ eV, while two other values are smaller than $\Delta_{\rm tr}$ (but all the values are comparable to $\Delta_{\rm tr}$ and $B=0.06$ eV). This readily explains the fact that even for $B=0$ there is a finite weight of the $a_{1g}^{\phantom{\pi}}$ states in the occupied part. Nevertheless, this weight clearly decreases with the decrease of $T$ and practically vanishes for $T=$ $120$ K. Therefore, it can be attributed to the finite temperature effects. On the other hand, when $B$ is finite, the same weight of the occupied $a_{1g}^{\phantom{\pi}}$ states practically does not depend on $T$, meaning that in this case it this the feature of Hund's second rule. 

\subsection{\label{sec:oo} Orbital ordering and magnetic ground state}
\par In order to study the orbital ordering in VI$_3$, we turn to the HF calculations. The use of the QE set of model parameters was important at the level of DMFT calculations to reproduce the semiconducting character of VI$_3$. However, the HF approximation tends to overestimate the band gap in comparison with DMFT (see Fig.~\ref{fig.dos}). To be specific, the energy gap in DMFT is well consistent with the experimental value of $0.6$ eV~\cite{VI3Kong}, while the HF band gap, $0.9$ eV, is clearly overestimated. Therefore, we believe that at the level of HF calculations, it would be more reasonable to use a smaller value of $U$ in order to mimic the band gap obtained in DMFT. This is the main reason why we switch to the LMTO set of model parameters (Table~\ref{tab:Coulomb}). The corresponding electronic structure can be found in Supplemental Material~\cite{SM}.

\par
\begin{figure}[t]
\begin{center}
\includegraphics[width=8.6cm]{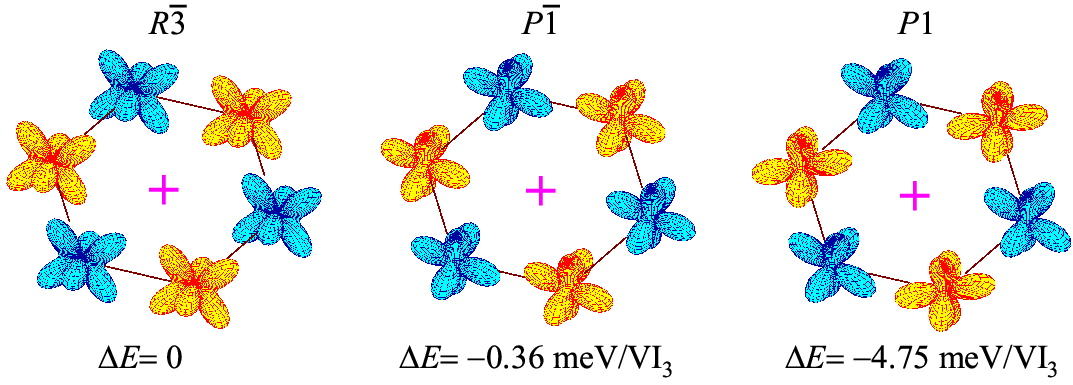} 
\end{center}
\caption{
Orbital ordering in the ferromagnetic state of rhombohedral $R\overline{3}$ phase of VI$_3$ as obtained in the Hartree-Fock calculations by enforcing the original trigonal $R\overline{3}$ symmetry, the triclinic $P\overline{1}$ symmetry, and fully relaxing the symmetry ($P1$). The crystallographic inversion centers are denoted by $+$. $\Delta E$ is the corresponding energy change relative to the ferromagnetic $R\overline{3}$ state.}
\label{fig.oo}
\end{figure}

 The orbital ordering (the distribution of electron densities around V sites) obtained in the HF approach for the FM state is displayed in Fig.~\ref{fig.oo}.
\noindent
\noindent Particularly, we have performed three types of calculations (by properly averaging the HF potential in the process of calculations): (i) by enforcing the original $R\overline{3}$ symmetry of the lattice, including threefold rotation about $z$ and inversion symmetries; (ii) by enforcing only the inversion symmetry and treating two V sites in the unit cell as equivalent (the space group $P\overline{1}$); (iii) by fully relaxing the symmetry and allowing different shape of the electron density around two V sites, which are crystallographically connected by the spatial inversion (the space group $P1$). In the $R\overline{3}$ case, both electrons are forced to occupy two $e_{g}^{\pi}$ states, resulting in the $e_{g}^{\pi} e_{g}^{\pi}$ configuration with small admixture of the $e_{g}^{\sigma}$ states due to the hybridization effects. In the $P\overline{1}$ case, one of the occupied states is $e_{g}^{\pi}$ while another one is $a_{1g}^{\phantom{\pi}}$ (the configuration $e_{g}^{\pi} a_{1g}^{\phantom{\pi}}$). Nevertheless, the occupied $e_{g}^{\pi}$ states are the same at the sites V1 and V2 (see Fig.~\ref{fig.LDA}). Thus, the threefold rotational symmetry is broken, but the spatial inversion is preserved. In the $P1$ case, the occupied $e_{g}^{\pi}$ states are different at the sites V1 and V2, thus breaking both threefold rotational and inversion symmetries. The total energy steadily decreases in the direction $R\overline{3} \to P\overline{1} \to P1$. Thus, the $R\overline{3}$ phase experiences the internal instability due to the orbital ordering, which tends to low the symmetry.

\par By enforcing $B=0$ in the HF calculations, we were able to obtain only one solution corresponding to the $R\overline{3}$ symmetry. Other solutions with the $P\overline{1}$ and $P1$ symmetries, obtained for $B \ne 0$, steadily converge to the $R\overline{3}$ one after setting $B = 0$. Thus, the Hund's second rule effects play a crucial role in breaking the symmetry and establishing the antiferro orbital order in the $P1$ phase. Yet, the precise meaning of ``antiferro orbital order'' in this context needs to be clarified, because the occupied $e_{g}^{\pi}$ orbitals in the vanadium sublattices V1 and V2 are not orthogonal and, strictly speaking, there are simultaneously ferro and antiferro components of the orbital ordering. Nevertheless, contrary to the $P\overline{1}$ phase, where there is only ferro component, the $P1$ phase also contains the antiferro one. 

\par
\begin{figure}[t]
\begin{center}
\includegraphics[width=8.6cm]{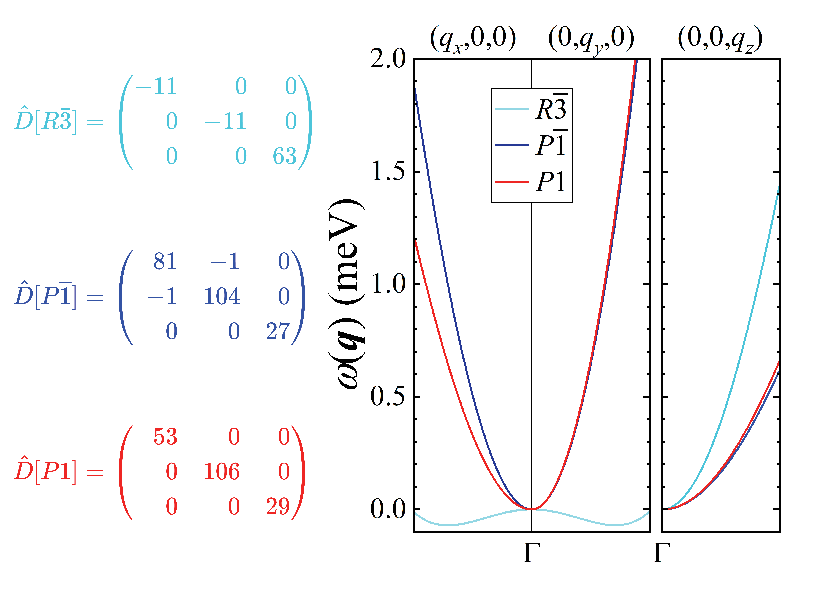} 
\end{center}
\caption{
Spin-wave stiffness tensors (in meV\AA$^2$) for the orbital states of the $R\overline{3}$, $P\overline{1}$, and $P1$ symmetry, and corresponding spin-wave dispersions near $\Gamma$-point of the Brillouin zone.}
\label{fig.sw}
\end{figure}

 Employing linear response theory~\cite{review2024}, we study the local stability of the FM state. The spin-wave stiffness tensors, $\hat{D}$, calculated for each of the orbital states of the $R\overline{3}$, $P\overline{1}$, and $P1$ symmetry, and the corresponding spin-wave dispersions near the $\Gamma$-point of the Brillouin zone are displayed in Fig.~\ref{fig.sw}.
\noindent
\noindent For the $R\overline{3}$ symmetry, the tensor $\hat{D}$ is negative-definite in the $xy$ plane, meaning that the FM state is unstable. With lowering the symmetry $R\overline{3} \to P\overline{1} \to P1$, the tensor $\hat{D}$ becomes positive-definite. Thus, the FM order is stabilized by the orbital order in the $P\overline{1}$ and $P1$ phases. Furthermore, lowering the symmetry results in the anisotropy of $\hat{D}$ in the $xy$ plane, which increases in the direction $P\overline{1} \to P1$. The AFM alignment of two V sublattices was also considered. However, the obtained AFM phases were substantially higher in energy than in the FM ones. The details can be found in Supplemental Material~\cite{SM}.

\par We also tried to choose various starting conditions in the HF calculations, assuming different populations of the atomic states in the initial guess for the potential (\ref{eqn.HF}) and then solving the problem self-consistently. Nevertheless, such calculations steadily converged to one of the orbital ordering displayed in Fig.~\ref{fig.oo}. Thus, the degeneracy of the orbitally ordered states, which was encountered, for instance, in the toy-model analysis in Sec.~\ref{sec:honeycomb}, is lifted, even without lattice distortions. The reason is that the degeneracy of unoccupied $e_{g}^{\sigma}$ states is lifted by electron-electron interactions with the occupied configuration $e_{g}^{\pi} a_{1g}^{\phantom{\pi}}$ and one of the $e_{g}^{\pi}$ states. Therefore, the virtual hoppings into the subspace of unoccupied $e_{g}^{\sigma}$ orbitals, relevant to the SE process, are no longer equivalent, which pick up the specific type of the orbital ordering. In the $R\overline{3}$ crystal structure, the only equivalent types of the orbital order can be obtained by rotating the ones in Fig.~\ref{fig.oo} by $\pm 120^\circ$ about $z$. Then, the orbital ordering pattern of the $R\overline{3}$ symmetry will transform to itself, while for each of the $P\overline{1}$ and $P1$ symmetries, such rotations will generate two more equivalent domains. Then, applying the inversion operation to such three orbital domains of the $P1$ symmetry, one can generate three more domains.

\par
\begin{figure}[t]
\begin{center}
\includegraphics[width=8.6cm]{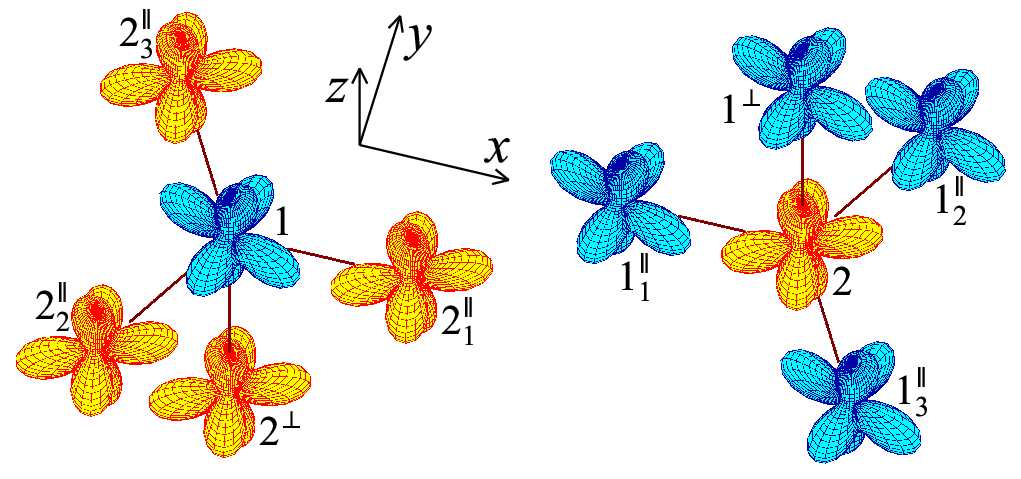} 
\end{center}
\caption{
Orbital ordering of the $P1$ symmetry around two V sites in the rhombohedral cell.}
\label{fig.oo2}
\end{figure}

 In the $P1$ state, the orbital ordering breaks the inversion symmetry and induces the electric polarization $\boldsymbol{P}$. The latter can be evaluated using Berry-phase theory~\cite{FE_theory1,FE_theory2,FE_theory2}, which can be adapted for the model Hamiltonian (\ref{eqn.ManyBodyH}) in the HF approximation~\cite{PRB2019}. Without SO interaction, it yields $\boldsymbol{P}= (-0.06,0.18,-0.05)$ $\mu {\rm C}/{\rm cm}^2$. Thus, $\boldsymbol{P}$ appears to be finite in the honeycomb plane as well as in the direction perpendicular to the plane. In the SE approximation, which can be derived starting from the general Berry-phase theory~\cite{PRB2020,PRB2019}, this polarization takes the pairwise form $\boldsymbol{P} = \sum_{\langle ij \rangle} \boldsymbol{P}_{ij}$ and can be expressed via weights of the Wannier functions, transferred to the neighboring sites, $| \alpha_{i \to j} |^2$: 
\noindent
\begin{equation}
\boldsymbol{P}_{ij} = \frac{e  \boldsymbol{\tau}_{ij}}{V} \left( | \alpha_{j \to i} |^2  - | \alpha_{i \to j} |^2 \right),
\label{eqn.PSE}
\end{equation}
\noindent where $\boldsymbol{\tau}_{ij} = \boldsymbol{R}_{i}$$-$$\boldsymbol{R}_{j}$  is the vector connecting the atomic site $j$ with the site $i$. The crystal structure of VI$_3$ is such that around each site V1 there are three neighboring sites V2 in the honeycomb plane ($2_{1}^{\parallel}$ - $2_{3}^{\parallel}$ in Fig.~\ref{fig.oo2}, separated by $3.95$~\AA) and one next-nearest-neighbor site V2 in the perpendicular direction ($2_{\phantom{1}}^{\perp}$, separated by $6.55$~\AA).
\noindent
\noindent In the $R\overline{3}$ structure, the sublattices V1 and V2 are transformed to each other by the special inversion. Therefore, such bonds are centrosymmetric and $| \alpha_{i \to j} |^2  = | \alpha_{j \to i} |^2$. Nevertheless, the orbital order breaks the inversion symmetry, so that besides the symmetric part, each of $| \alpha_{i \to j} |^2$ will also acquire the antisymmetric one, resulting in finite $\boldsymbol{P}_{ij}$. Similar situation occurs around site $2$. If for each of the bond $1j$ around site $1$, $2j'$ is the equivalent to it bond around site $2$ (say, $12_{1}^{\parallel}$ and $21_{1}^{\parallel}$, or $12_{\phantom{1}}^{\perp}$ and $21_{\phantom{1}}^{\perp}$ in Fig.~\ref{fig.oo2}), we have $| \alpha_{1 \to j} |^2 = | \alpha_{j' \to 2} |^2$, $\boldsymbol{\tau}_{1j} = -\boldsymbol{\tau}_{2j'}$, and therefore $\boldsymbol{P}_{1j} = \boldsymbol{P}_{2j'}$, resulting in final total $\boldsymbol{P}$. Thus, the bonds $12_{\phantom{1}}^{\perp}$ and $21_{\phantom{1}}^{\perp}$ are responsible for finite $P^{z}$, while other bonds are responsible for finite polarization in the honeycomb plane.

\subsection{\label{sec:tr} Triclinic distortion}
\par In this section we briefly consider the effect of triclinic distortion on the orbital ordering, using for these purposes experimental parameters of the $P\overline{1}$ structure at $T=9$ K~\cite{structure}. The triclinic distortion additionally splits the $e_{g}^{\pi}$ levels. However, the magnitude of this splitting ($\sim 10$ - $20$ meV~\cite{SM}) is smaller than $B$, so that the main tendencies obtained for the trigonal structure remain largely intact. 

\par
\begin{figure}[t]
\begin{center}
\includegraphics[width=8.6cm]{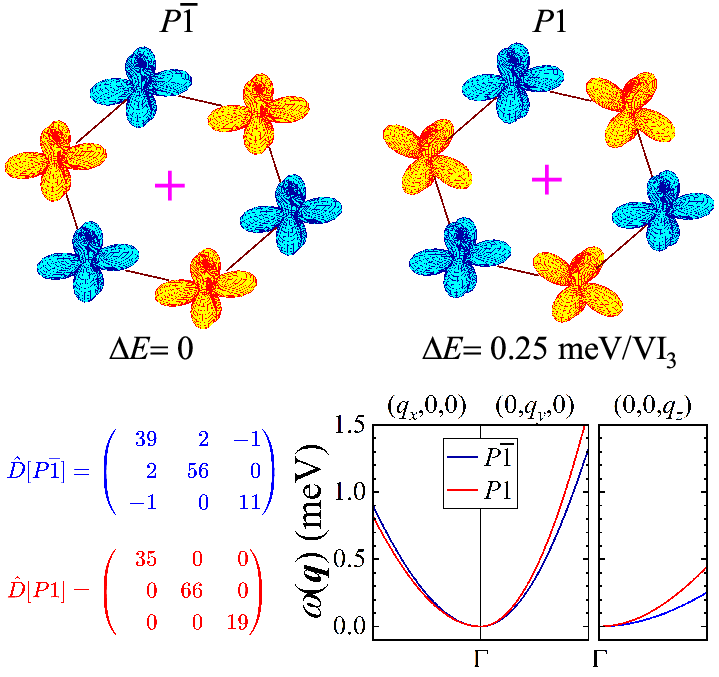} 
\end{center}
\caption{
Top: Two types of the orbital ordering of the $P\overline{1}$ and $P1$ symmetry, obtained in the triclinic structure of VI$_3$ for the ferromagnetic state. The crystallographic inversion center is denoted by $+$. $\Delta E$ is the energy change relative to the orbital state of the $P\overline{1}$ symmetry. Bottom: Spin-wave stiffness tensors (in meV\AA$^2$) for the orbital states of the $P\overline{1}$ and $P1$ symmetry, and corresponding spin-wave dispersions near $\Gamma$-point of the Brillouin zone.}
\label{fig.ootr}
\end{figure}

 In the HF approximation, we were able to obtain two distinct solutions for the FM state (Fig.~\ref{fig.ootr}).
\noindent
\noindent The first one has $P\overline{1}$ symmetry, the same as the crystal structure, where two V sublattices are transformed to each other by the spatial inversion. The second solution has $P1$ symmetry, where the inversion symmetry is broken by the antiferro orbital order. The solutions are nearly degenerate. The spin-wave stiffness tensor $\hat{D}$ is positive-definite for both symmetries, meaning that the FM state is stable. Enforcing $B=0$ in the HF calculations, the $P1$ solution steadily relaxes to the $P\overline{1}$ one. Thus, the Racah parameter $B$ is solely responsible for the antiferro orbital order and breaking the inversion symmetry.

\subsection{\label{sec:SO} Spin-orbit interaction and magnetic-field control of electric polarization}
\par Similar to more studied CrI$_3$~\cite{Lado}, the main source of the SO interaction in VI$_3$ are the heavy iodine atoms. The corresponding parameter of the SO coupling, $\xi \sim 0.1$ eV, is comparable to $B$ and larger than $\Delta_{\rm tr}$. Moreover, unlike in CrI$_3$, the majority-spin $t_{2g}$ shell in VI$_3$ is only partially filled. Therefore, one can generally expect the SO coupling to play a very important role in VI$_3$.

\par
\begin{figure}[t]
\begin{center}
\includegraphics[width=8.6cm]{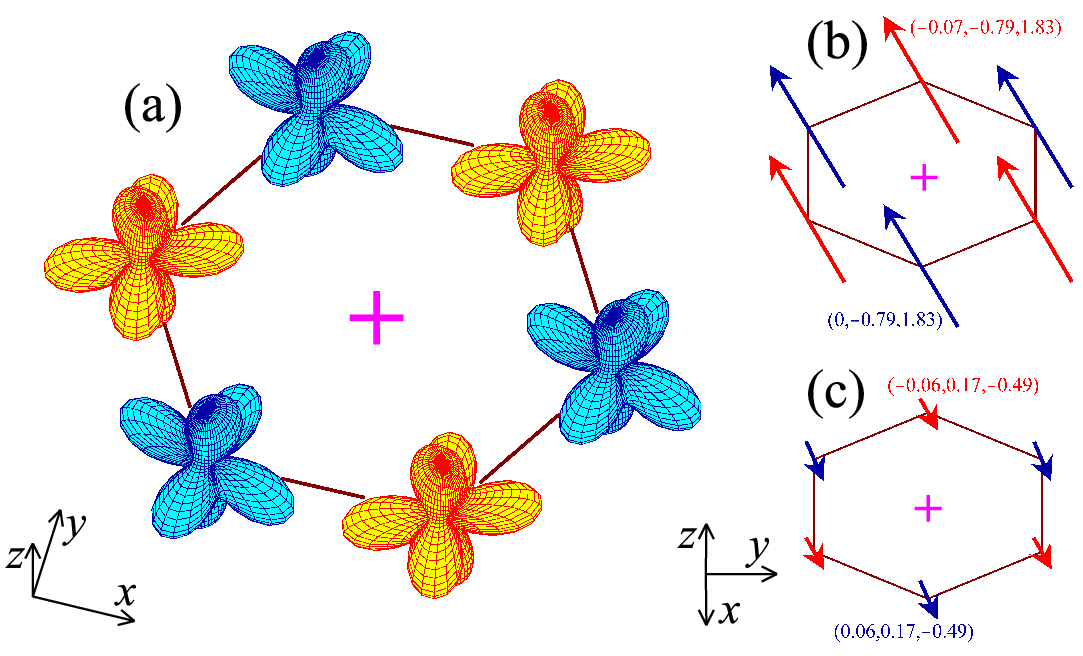} 
\end{center}
\caption{
Results of Hartree-Fock calculations with the spin-orbit interaction: (a) Orbital ordering of the $P1$ symmetry, (b) Spin magnetic moments ($\boldsymbol{M}_{\rm S}^{\nu}$), (c) Orbital magnetic moments ($\boldsymbol{M}_{\rm L}^{\nu}$). The numerical values of $\boldsymbol{M}_{\rm S}^{\nu}$ and $\boldsymbol{M}_{\rm L}^{\nu}$ in two magnetic sublattices are given in  parentheses. The crystallographic inversion center is denoted by $+$. }
\label{fig.oorr}
\end{figure}

 The SO interaction practically does not change the distribution of electron density around V sites (see Fig.~\ref{fig.oorr}a). 
\noindent
\noindent However, it has a profound effect on other elements of the density matrix, responsible for magnetic properties. Furthermore, besides spin, there is an appreciable orbital magnetization. Breaking the threefold rotation symmetry in the orbitally ordered states of the $P\overline{1}$ and $P1$ symmetry will lead to a canted magnetic structure. On the microscopic level, this canting is related to the population of the $a_{1g}^{\phantom{\pi}}$ orbital and one of the $e_{g}^{\pi}$ orbitals, so that the SO interaction operating between the occupied $a_{1g}^{\phantom{\pi}}$ and unoccupied $e_{g}^{\pi}$ orbitals will obey the selection rules: $\Delta \sigma = \pm 1$ and $\Delta m = \mp 1$, leading to rotation of spin ($\boldsymbol{M}_{\rm S}^{\nu}$) and orbital ($\boldsymbol{M}_{\rm L}^{\nu}$) magnetic moments away from the $z$ axis. In the $P\overline{1}$ state, where two V sites remain equivalent, this rotation occurs in the same direction for $\nu=$ V1 and V2. Thus, the canting is ferromagnetic. The additional breaking of inversion symmetry in the $P1$ state will lead to the AFM canting of spin and orbital magnetic moments, which can be regarded as an effect of Dzyaloshinskii-Moriya interaction~\cite{Dzyaloshinskii_weakF,Moriya_weakF}, induced by the antiferro orbital ordering in the otherwise centrosymmetric crystal structure. Microscopically, the AFM canting occurs because, in the $P1$ state, the unoccupied $e_g^{\pi}$ orbital at the sites $1$ and $2$ are different, resulting in the different coupling with the occupied $a_{1g}^{\phantom{\pi}}$ orbitals. The FM canting takes place mainly in the $yz$ plane (see Figs.~\ref{fig.oorr}b and \ref{fig.oorr}c). First, we note that beside spin, there is a large orbital magnetic moment $M_{\rm L}^{z} = 0.5$ $\mu_{\rm B}$, which is consistent with the experimental value of $0.6$ $\mu_{\rm B}$ derived from the x-ray magnetic circular dichroism spectra~\cite{MLexp}. The polar angles formed by the spin, orbital, and total ($\boldsymbol{M}_{\rm S}^{\nu}$$+$$\boldsymbol{M}_{\rm L}^{\nu}$) magnetic moments with the $z$ axis can be estimated as $\vartheta_{\rm S}=24^{\circ}$, $\vartheta_{\rm L}=160^{\circ}$, and $\vartheta_{\rm S+L}=25^{\circ}$, respectively. The latter is consistent with the experimental estimate of $\vartheta_{\rm S+L}^{\rm exp} \sim 36^{\circ}$~\cite{VI3Hao}. Then, the AFM canting takes place mainly in the $xy$ plane, where the magnetic moments of the sublattices V1 and V2 are additionally rotated relative to each other by $\Delta \varphi_{\rm S} =5^{\circ}$ and $\Delta \varphi_{\rm L} =36^{\circ}$, for the spin and orbital counterparts, respectively. 

\par Thus, the SO interactions largely modifies the magnetic structure of VI$_3$. This changes the electric polarization dramatically. Using Berry-phase theory, one can readily evaluate the electronic part of $\boldsymbol{P}$, associated with the change of the electronic structure after taking into account the SO coupling. It yields $\boldsymbol{P}= (-1.86, 1.59,-3.85)$ $\mu {\rm C}/{\rm cm}^2$, which exceeds the same value obtained without SO coupling by more than one order of magnitude. This change is associated with the additional contributions to the magnetoelectric coupling, which are activated by the SO interaction. First, the orbital magnetization can additionally contribute to $\boldsymbol{P}$~\cite{Malashevich2012,Scaramucci}, which is a quite plausible scenario in the present case because $\boldsymbol{M}_{\rm L}$ is large. Another contribution to $\boldsymbol{P}$ is due to the noncollinear magnetic alignment~\cite{KNB,PRB2017,PRL2021}. 

\par
\begin{figure}[b]
\begin{center}
\includegraphics[width=8.6cm]{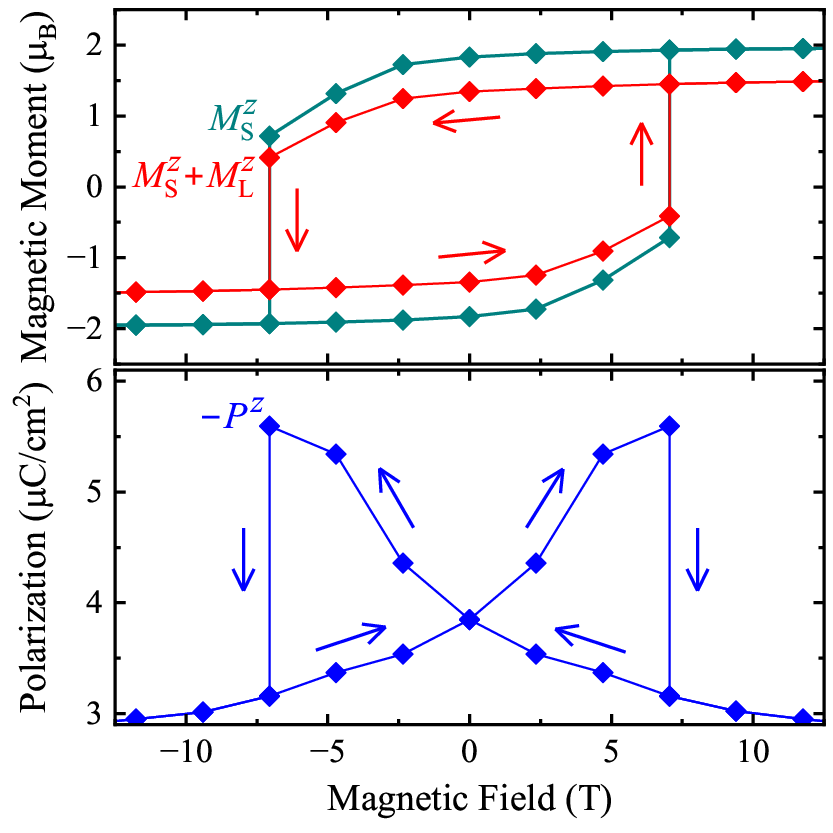} 
\end{center}
\caption{
Magnetic-field dependence of magnetization (top) and electric polarization (bottom). The field is applied parallel to the $z$ axis. $M^{z}_{\rm S}$ and $M^{z}_{\rm S}$$+$$M^{z}_{\rm L}$ are the $z$-components of, respectively, spin and total magnetic moments. $P^{z}$ is the $z$-component of the electric polarization.}
\label{fig.field}
\end{figure}

 Then, we consider how the electric polarization can be controlled by external magnetic field $\boldsymbol{H}$. The basic idea is that by applying the magnetic field one can control the canting of magnetic moments and the degree of  mixing of the $a_{1g}^{\phantom{\pi}}$ and $e_{g}^{\pi}$ characters in the ground state, which plays a crucial role in establishing the antiferro orbital ordering and developing the electric polarization. Thus, $\boldsymbol{P}$ can be eventually controlled by $\boldsymbol{H}$. Then, we apply the magnetic field parallel to $z$, $\boldsymbol{H} = (0,0,H)$, and monitor the behavior of spin and orbital magnetic moments as well as the electric polarization, derived from the HF calculations. The results are summarized in Fig.~\ref{fig.field}: the magnetization reveals the specific hysteresis loop, while the polarization has a butterfly-like shape. 
\noindent
\noindent The magnetic field $H \sim 10$ T applied in the direction of $M^{z}_{\rm S}$ is sufficient to saturate both magnetization and the electric polarization ($P^{z} \sim 3$ $\mu {\rm C}/{\rm cm}^2$). The magnetic field in the opposite direction gradually decreases $M^{z}_{\rm S}$ and $M^{z}_{\rm S}$$+$$M^{z}_{\rm L}$ and increases the $xy$-components of these moments. Then, $H \sim -7$ T causes the reorientation of $M^{z}_{\rm S}$ along the field. The corresponding polarization undergoes the jump $\Delta P^{z} \sim 2.4$ $\mu {\rm C}/{\rm cm}^2$, which is comparable to $\Delta P^{z} \sim 1.7$ $\mu {\rm C}/{\rm cm}^2$ in CaBaCo$_4$O$_7$ and so far regarded as the largest experimentally observed change of  electric polarization induced by the magnetic field~\cite{CaBaCo4O7}.

\section{\label{sec:Summary} Summary and Outlook}
\par We have proposed a new route for designing ferroelectric ferromagnets -- a fundamentally and practically important subclass of multiferroic materials, which are not simply magnetic, but ferromagnetic. Our basic idea is that the antiferro orbital ordering across the inversion center should not only produce the FM interactions between the spins, as it follows from the GKA rules, but can also break the inversion symmetry. The vitality of this idea was illustrated on the toy models of the orbital ordering in the zigzag chain and honeycomb plane. Then, we have proposed that such scenario can be indeed realized in the van der Waals ferromagnet VI$_3$, where the Hund's second rule effects tend to form the atomic ground state with the greatest possible multiplicity, thus unquenching the orbital degrees of freedom and activating the KK mechanism of the orbital ordering. This mechanism is responsible for the antiferro orbital order in VI$_3$, which breaks not only the threefold rotation, but also inversion symmetry, resulting in the FE-FM ground state. Thus, the orbital degeneracy is lifted, as it is required by Jahn-Teller theorem~\cite{JahnTeller}. However, this degeneracy lifting occurs via the SE processes, whereas the crystal distortions probably play a secondary role. This is in line with general symmetry considerations suggesting that the pseudo Jahn-Teller mechanism, which results in the noncentrosymmetric ionic displacements, is not operative in the $d^{2}$ systems, such as VI$_3$~\cite{Bersuker2012}. 

\par The relativistic SO interaction, collaborating with the symmetry breaking, results in the canting of magnetization, which can be further manipulated by the magnetic field. This opens a possibility for controlling the electric polarization, which undergoes a huge change in the magnetic field.

\par The available experimental information about the crystal structure of VI$_3$, especially regarding the stacking of the honeycomb planes as well as the symmetry of these planes, is very controversial, as several different structures have been proposed for the room-temperature as well as low-temperature phases~\cite{VI3Tian,VI3Dolezal,VI3Son,VI3Gati,structure,VI3Hao}. We consider such fragility of the crystal structure to be a manifestation of the orbital phenomena in VI$_3$: it appears that the orbital degrees of freedom in VI$_3$ remain flexible and there may be several scenarios of lifting the orbital degeneracy depending on the experimental conditions. We hope that our scenario, where the orbital ordering not only lifts the degeneracy but also breaks the inversion symmetry, leading to the ferroelectric ferromagnetism, can be eventually realized in VI$_3$. From this perspective, particularly interesting are the results of Ref.~\cite{VI3Son}, where the $P\overline{3}1c$ and $C2/c$ structures were proposed for, respectively, the high- and low-temperature phases of VI$_3$. These structures are still centrosymmetric. However, the vanadium sites V1 and V2, forming the honeycomb planes, become inequivalent. Thus, within each honeycomb plane, the inversion symmetry appears to be broken and this is consistent with our scenario of the antiferro orbital order. Another, again indirect, indication of the inversion symmetry breaking in the honeycomb plane is the different behavior of the magnetic sublattices V1 and V2 reported in certain temperature range ($36$ K $<T<$ $51$ K)~\cite{VI3Gati}.

\par There is a number of theoretical studies reporting the threefold rotation breaking in VI$_3$ at the level of DFT$+$$U$ calculations with the SO coupling, due to partial population of the $a_{1g}^{\phantom{\pi}}$ state~\cite{MLexp,Yang,Sandratskii,Nguyen,Camerano}. Nevertheless, none of these studies reported the inversion symmetry breaking. The symmetry breaking in Refs.~\cite{MLexp,Yang,Sandratskii,Nguyen,Camerano} is solely related to the SO interaction: the $e_{g}^{\pi} e_{g}^{\pi}$ configuration, respecting the threefold rotation symmetry, would yield only small $\boldsymbol{M}_{\rm L}$ (being about $-$$0.1$ $\mu_{\rm B}$, along the $z$ axis, emerging due to the mixing of $e_{g}^{\pi}$ with the unoccupied $e_{g}^{\sigma}$ states by the SO coupling). Therefore, in order to increase $\boldsymbol{M}_{\rm L}$ and thus maximize the energy gain caused by the SO interaction, it is essential to rotate $\boldsymbol{M}_{\rm L}$ away from the $z$ axis by breaking the threefold rotation symmetry and populating the $a_{1g}^{\phantom{\pi}}$ state. What we propose here is fundamentally different: according to our scenario, the symmetry can be broken by ordering the orbitals, which would remain degenerate in the atomic limit. Formally, no SO interaction is needed in our case, though it plays an important role by further facilitating the symmetry breaking and for establishing the magnetic-field control of $\boldsymbol{P}$. Whether such behavior can be indeed achieved at the level of DFT$+$$U$ calculations depends on the implementation, which must include all necessary terms proportional to the Racah parameter $B$. Although it was considered on earlier stages, where the DFT$+$$U$ functional was formulated in terms of all Slater integrals and, therefore, explicitly included the dependence on the Racah parameter $B$~\cite{PRB1994,Liechtenstein1995}, the later, commonly used but simplified, versions are formulated in terms of only one parameter $U_{\rm eff} = U-J$ and, thus, disregard the contributions responsible for Hund's second rule~\cite{PRB1996,Dudarev}.

\par In the most general formulation, DFT should be able to incorporate the exchange-correlation interactions responsible for atomic Hund's second rule. However, such effects are omitted in many popular approximations supplementing DFT, such as LDA or GGA, which take the functional form of these interactions from the limit of homogeneous electron gas, where, strictly speaking, the atomic Hund's rules are no longer applicable, resulting in a number of fundamental issues for atomic systems~\cite{Weinert}. Therefore, a very popular direction around 1990s was to simulate the Hund's rule physics on the top of LDA by introducing a phenomenological correction, proportional to some appropriate Racah parameter ($B$ for $3d$ electrons), with the aim to reproduce the orbital magnetization in solids, which was severely underestimated in the local spin-density approximation~\cite{Brooks1989,Norman1990,Norman1991}. For the Mott insulators, the logic behind was that the Coulomb repulsion $U$ should split the occupied and unoccupied $3d$ states~\cite{Terakura1984}. However, the Racah parameter $B$ is another important ingredient to decide the correct symmetry of states, which will be further split by $U$~\cite{Norman1990}. However, in most of the cases, such symmetry is already decided by the crystal field and SO interaction. Moreover, the orbital magnetization strongly depends on the value of Coulomb repulsion $U$, as it controls the strength of the hybridization between occupied and unoccupied states~\cite{PRL1998,PRL2005}. In many cases, the value of orbital magnetization can be reproduced by the Coulomb $U$ alone, especially when it is treated as an adjustable parameter. From this perspective, the unique aspect of VI$_3$ is that both parameters, $U$ and $B$, appear to be important for finding the correct ground state. 

\par While the importance of intraatomic exchange coupling $J$ in the physics of strongly correlated materials is well recognized today~\cite{Georges}, the more delicate effects, driven by the Racah parameter $B$, remain largely unexplored. It is true that $B$ is much smaller than $J$ (typically, $B \sim 0.1 J$), reflecting the well know hierarchy of atomic Hund's rules, when the second rule always follow the first one. Nevertheless, if $B$ is larger or comparable to the characteristic crystal field, the Hund's second rule effects can lead to a number of interesting and so far unexplored effects. The ferromagnetic ferroelectricity in VI$_3$, which we propose in this work, is certainly one of them. 

\section*{Acknowledgement}
\par MANA is supported by World Premier International Research Center Initiative (WPI), MEXT, Japan.

\

\par I.V.S. conceptualized the work, performed most of the calculations (except specified below), and wrote the manuscript. R.O. and S.A.N. performed the electronic structure calculations and constructed the model Hamiltonian using the QE method. S.A.N. performed the DMFT calculations. All authors discussed the results and commented on the manuscript.

\

\par The authors declare no competing interests.

\end{document}